# IoT Malware Detection Architecture using a Novel Channel Boosted and Squeezed CNN

Muhammad Asam[1, 2], Saddam Hussain Khan[1, 2], Tauseef Jamal[3], Asifullah Khan[1, 2, 4],

asim2k994@gmail.com, hengrshkhan822@gmail.com, jamal@pieas.edu.pk, asif@pieas.edu.pk

[1]Pattern Recognition Lab, Department of Computer & Information Sciences, Pakistan Institute of Engineering & Applied Sciences, Nilore, Islamabad 45650, Pakistan.
[2]PIEAS Artificial Intelligence Center (PAIC), Pakistan Institute of Engineering & Applied Sciences, Nilore, Islamabad 45650, Pakistan.
[3]Department of Computer & Information Sciences, Pakistan Institute of Engineering & Applied Sciences, Nilore, Islamabad 45650, Pakistan.
[4]Center for Mathematical Sciences, Pakistan Institute of Engineering & Applied Sciences, Nilore, Islamabad 45650, Pakistan.

`

*Abstract—*

Interaction between devices, people, and the Internet has given birth to a new digital communication model, the Internet of Things (IoT). The seamless network of these smart devices is the core of this IoT model. However, on the other hand, integrating smart devices to constitute a network introduces many security challenges. These connected devices have created a security blind spot, where cybercriminals can easily launch an attack to compromise the devices using malware proliferation techniques. Therefore, malware detection is considered a lifeline for the survival of IoT devices against cyberattacks. This study proposes a novel IoT Malware Detection Architecture (iMDA) using squeezing and boosting dilated convolutional neural network (CNN). The proposed architecture exploits the concepts of edge and smoothing, multi-path dilated convolutional operations, channel squeezing, and boosting in CNN. Edge and smoothing operations are employed with split-transform-merge (STM) blocks to extract local structure and minor contrast variation in the malware images. STM blocks performed multi-path dilated convolutional operations, which helped recognize the global structure of malware patterns. Additionally, channel squeezing and merging helped to get the prominent reduced and diverse feature maps, respectively. Channel squeezing and boosting are applied with the help of STM block at the initial, middle and final levels to capture the texture variation along with the depth for the sake of malware pattern hunting. The proposed architecture has shown substantial performance compared with the customized CNN models. The proposed iMDA has achieved Accuracy: 97.93%, F1-Score: 0.9394, Precision: 0.9864, MCC: 0. 8796, Recall: 0.8873, AUC-PR: 0.9689 and AUC-ROC: 0.9938.

**Keywords**— IoT, Malware, Detection, Convolutional Neural Network, Deep Learning, Channel boosting, Channel squeezing, Split-Transform-Merge.

1. **INTRODUCTION**

The concept of transforming real-world objects into virtual objects emerged as the Internet of Things (IoT). Under this concept, intelligent objects and devices can share data and resources according to the situation and environment [1]. This web of interconnected devices plays a vital role in our daily lives, ranging from health, smart homes, education, and, especially, industry. Masses are becoming familiar with the deployment of these devices in the field of agriculture, for soil condition monitoring [2], healthcare and e-health applications [3]–[5], and military domains [6], as well. Deployments of these gadgets range from operational areas to critical infrastructure

`

services. Industry 4.0 exploited this concept to build the link between supply chain, industrial production, and end-users [7]. The IoT ecosystem used in industry, Industrial IoT (IIoT), undoubtedly contributes towards the productivity and the quality of the industrial infrastructures.

IoTs lack secure design rules, and hence these have become an accessible playground for cybercriminals [8]. IoT devices are resource-constrained. These devices are usually installed with a default username password. Due to the embedded nature of the IoT devices, they are not patched regularly [9]. Network vulnerabilities for communicating with these devices can also be exploited easily at IoT touchpoints. The security protocol cannot be uniformly implemented on all the devices. The manufacturing of the devices does not conform to some consistent standards. These security challenges are depicted in Figure 1.

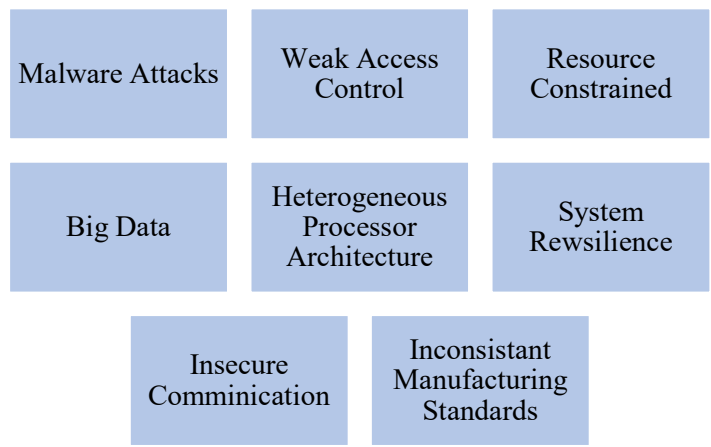

**Figure 1.** IoT Security Challenges

There are many heterogenic device structures and network protocols. They also possess a unique characteristic of processor heterogeneity [10]. So, the IoT industry lags unified security protocols for design and implementation. These weaknesses of IoT design enlarge the attack surface area and lead to security breaches. Cybercriminals utilize these attack surfaces for their illegal actions and exploit the vulnerabilities. Major cyber security concerns are host and Network Intrusions, malware attacks, compromised nodes, botnets, rootkits, ransomware, and DDoS. Therefore, a robust mechanism for detecting such activities is needed to detect and mitigate these digital security exploits quickly.

Research towards this security aspect of IoTs has attracted increased academic, industrial, and state-level attention. Several research efforts have discovered potential cyber threats and provided countermeasures against cyberattacks. Cyber security experts believe that most cyber exploits are carried out through malware attacks. Many research studies in the literature have

`

attempted this challenge of malware detection. Static, dynamic, hybrid, and image-based malware analysis comes under this challenge's broad categories [11].

Machine learning techniques have been extensively used for malware detection as they are more robust and give promising performance [12], [13]. Anti-malware tools have achieved improved performance with the help of machine learning tools. Several machine learning algorithms have been employed for mining the vulnerabilities in the IoT firmware and IoT applications that can infect and corrupt the edge devices and the whole network of the connected devices. Recent advancement in machine learning has proved its capabilities in detecting and classifying IoT malware. Research studies for anti-malware applications have increased the inclination towards machine learning tools and techniques. Computational power improvement has also enhanced the performance of machine learning strategies for malware detection and classification. Application of the machine learning needs the features of the IoT malware for making their verdict.

As the malware databases are increased, deep learning techniques suited more pertinent for the detection and analysis. Recent research has been molded towards the application of neural networks in the field of malware analysis. Neural networks, especially deep convolution neural networks (CNNs), have proven their competencies for feature extraction and feature identification in IoT malware. Deep CNNs build the malware detection systems by defining the discriminative features in IoT malware. Deep CNNs show enhanced performance as these models learn the complicated features of the IoT malware at different abstraction levels. Features learned in the lower layers are enriched in the upper lawyers. These features are extracted from the visual images of the problem domain.

IoT Malware dataset exploited in the current study has not been addressed previously to the best of our knowledge. This study utilized the image representation of IoTs malware and benign files. It is observed that deep CNN has shown promising performance for the visual challenges [14]. We have proposed applying deep learning techniques for the malware detection challenge.

Our main contributions in the current study are described below.

1. A novel IoT Malware detection architecture (iMDA), using squeezing and boosting dilated CNN, is proposed for IoT Malware analysis using a new benchmark dataset.
2. The proposed iMDA incorporates the edge and smoothing, multi-path dilated convolutional, channel squeezing, and boosting operations in CNN. Edge and smoothing

`

operations are employed within split-transform-merge (STM) blocks to extract local structure and minor contrast variation in the malware images.

3. STM blocks performed multi-path dilated convolutional operations, which helped to recognize the global structure of malware patterns. Additionally, channel squeezing and boosting are applied at different granular levels to get the reduced but prominent and diverse feature maps for capturing texture variations.

4. The proposed iMDA has shown significant performance compared with existing CNNs via TL in terms of standard performance metrics using MCC, F1-Score, AUC, Accuracy, Precision, and Recall.

The rest of the paper is structured as follows: the next section specifies related work in IoT malware analysis. Section 3 explains the proposed novel malware detection methodology. Section 4 describes the experimental setup. Section 5 discusses the results of our work. The conclusion is described in section 6.

## 2. Related Work

IoT Malware analysis is carried out using static, dynamic, and hybrid analysis techniques. Nataraj et al. [15] were the first to perform the malware analysis based on greyscale images in 2011. Malware visual images are created by transcribing the eight-bit code value of the executable files to the corresponding greyscale value. Image texture features are extracted from these images. The idea of texture-based analysis for IoT malware is emerging in context with deep learning. Evanson et al. [16] proposed an approach for malware analysis using texture images of malware files and machine learning in IoTPOT [17] for Bashlite and Mirai. They came up with the Haralick image texture features from gray-level co-occurrence matrix and used machine learning classifiers. Carrilo et al. [18] explored the malware forensic and reverse engineering capabilities for malware characterization. They first used machine learning for malware detection for Linux-based system malware of IoT. They also discovered new malware detection by using clustering techniques. They exploited the dataset provided by E. Cozzi et al. [19]. Ganesh et al. [20] exploited machine learning capabilities to detect Mirai botnet attacks in IoTs. They applied ANN to evaluate their approach on the N-BaIoT dataset. Bendiab et al. [21] applied deep learning for malware analysis traffic IoT. They applied ResNet50 for the experimental verification of their concept using a 1000 network (pcap) file.

`

Kyushu et al. [22] proposed a lightweight approach for IoT malware detection. They targeted the DDoS malware for their study and extracted the malware images from malware binaries in IoTPOT [17]. Their experimental setup showed performance for detecting the DDoS malware and good-ware. Ren et al. [23] gave an end-to-end malware detection mechanism for Android IoT devices. They collected 8000 benign and 8000 malicious APK files from the Google Play store and VirusShare, respectively. They used the significance of deep learning for the evaluation of their concept. Naeem et al. [24] ] detected the malware in Industrial IoT by proposing deep CNN-based traffic, behavior, and log databases analysis. They utilized the color images of the targeted malware for detection in the Leopard Mobile dataset.

However, the evaluation of the reported work is presented in Accuracy and Precision. Practically, malware datasets are imbalanced. Therefore, other evaluation metrics must be taken into account. In this regard, our proposed research work exploited the benchmark Kaggle IoT dataset. Performance evaluation metrics F1-Score, MCC, AUC-PR, and AUC-ROC are also evaluated along with Accuracy and Precision.

3. METHODOLOGY

3.1. *Data Augmentation*

CNN models give better generalization upon large labeled data. Sometimes, the data points for the model training are not adequate. The data augmentation technique produces the artificial sample points by applying image transformation operations [25]–[27]. These operation includes rotation (0-360 degree), scaling (0.5-1), shearing (-0.5, +0.5) and reflection (in left and right direction). The augmentation process helped to improve generalization and made the dataset more robust.

3.2. *Proposed IoT malware detection architecture (iMDA)*

This study proposes a novel image-based IoT malware detection architecture, iMDA. The suggested architecture discriminates the malware image sample from benign images. Spit-Transform-Merge (STM) is the main building block of this architecture. Three STM-based blocks concept is systematically implemented using region and edge detection operations.

The concept of channel boosting is imparted for high precision, improving the detection rate. Implementation details of the proposed architecture are highlighted in Figure 2. The performance of the proposed architecture is compared with the existing CNN models using TL-based implementation, as shown in Figure 3.

`

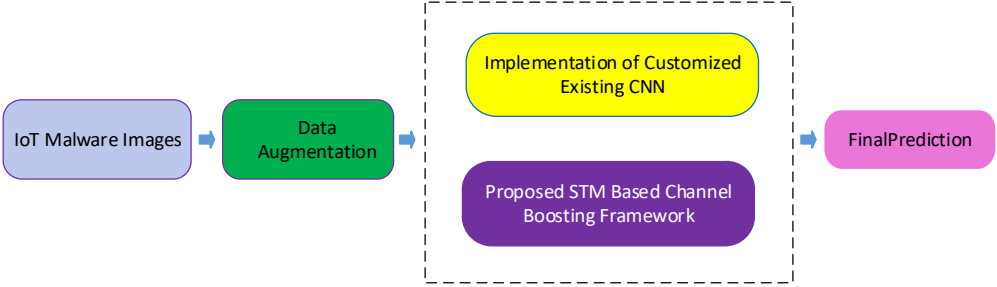

**Figure 2.** A Brief overview of the proposed framework.

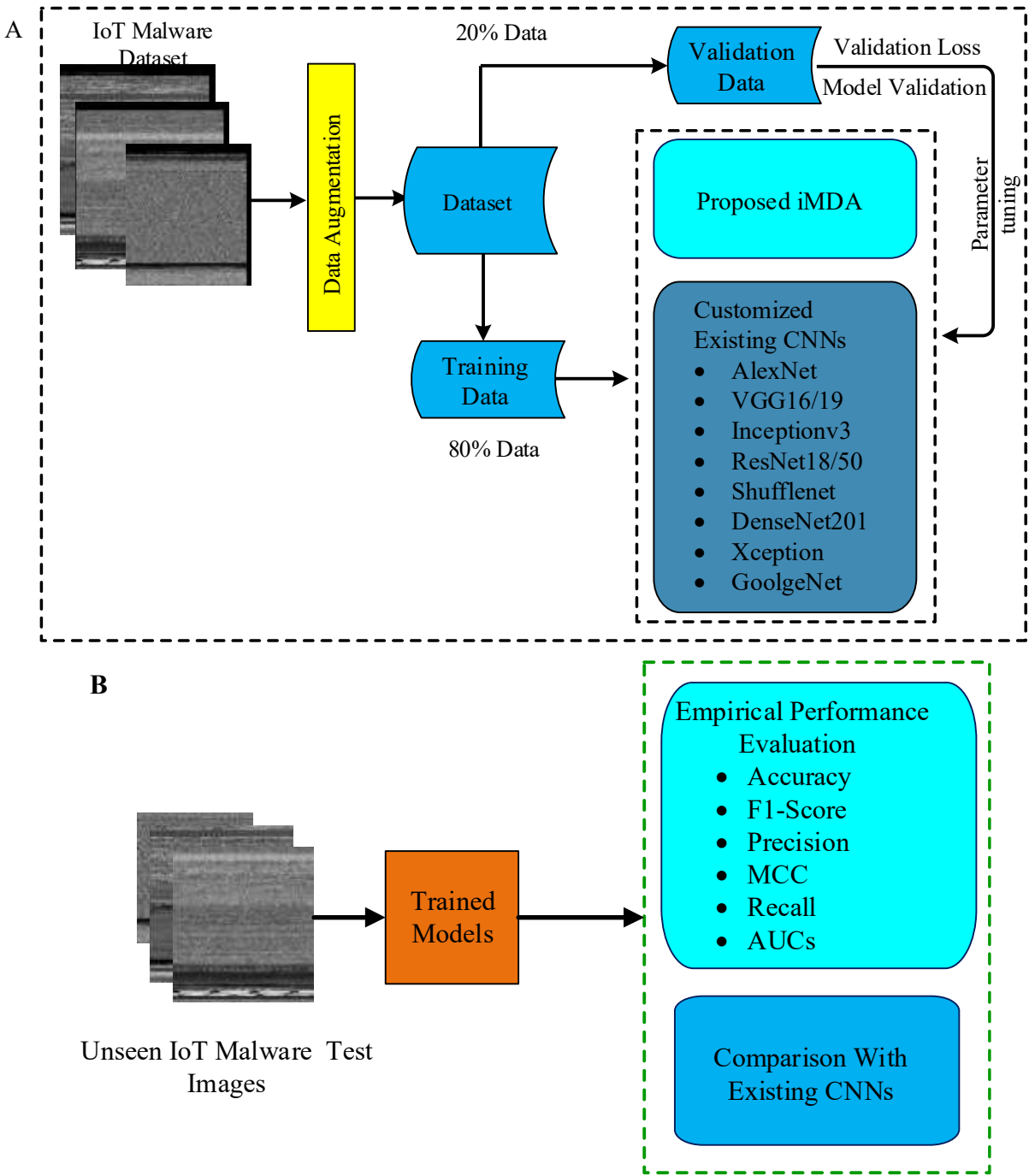

**Figure 3.** Detailed overview of Training (A) and Testing (B) of the proposed framework

## 3.3. *Proposed channel squeezing and boosting blocks*

Deep CNN models are powerful and robust for their texture feature mining abilities. These models use convolutional operations for exploiting structural information in the image data. These operations are used to extract the dataset's features according to the target domain. This innovative feature of the deep CNN is utilized in the current architecture for IoT malware detection. This architecture is tailored by proposing a concatenated STM-based channel boosting approach [28].

The proposed STM block comprises a stack of four blocks, as shown in Figure 4(b). Details of the operations performed in each block are shown in Figure 4(c). Block- B and block C employ the same convolutions, batch normalization, and Relu operation with max and average pooling operation. Two convolution operations employed in each block are used to extract the features information at the detailed and abstract levels, respectively. Block D and Block E employ the three convolution operation. Two operations are used for the detailed features extraction, while one is used for the abstract level feature information extraction. These convolution operations are followed by batch normalization, ReLu, and pooling operations. Average and max pooling operations are employed alternatively for edge and region-based feature extraction [29], [30]. Finally, the concept of channel boosting is used at the STM block exit to enhance the discrimination ability of the architecture.

The STM block splits the input IoT malware image data into four branches to feed the four blocks of the STM. These blocks learn the region and edge-based informative features at a different level of abstraction from the input dataset. This learning helps to gather the highly discriminating features of the IoT malware at a high and detailed level. This info is imparted into different channels from each block. Information infused in other channels is concatenated at the exit of the STM block. This channel boosted feature space is rich in diverse levels of textural feature information of the malware.

$$x_{a,b} = \sum_{i=1}^{p}\sum_{j=1}^{q} x_{a+i-1,b+j-1} * f_{i,j} \qquad (1)$$

$$(xavg)_{a,b} = \frac{1}{s^2}\sum_{i=1}^{s}\sum_{j=1}^{s} x_{a+i-1,b+j-1} \qquad (2)$$

$$(xmax)_{a,b} = max_{i=1,\ldots s, j=1,\ldots s} x_{a+i-1,\ b+j-1} \qquad (3)$$

$$C_{Boost} = merge(C_B||C_C||C_D||C_E) \qquad (4)$$

$$x = \sum_{d=1}^{D}\sum_{c=1}^{C} u_d x_c \qquad (5)$$

Eq. (1) shows the convolution operation of filter $f$ and input channel $x$ of size p × q and A × B, respectively. The dimensions of the convolved output range from 1 to A - p + 1 and B - q + 1, respectively. 's' denotes the dimension for average and max-pooling operations, shown Eq. (2) and Eq. (3). In Eq. (4), $C_B$, $C_C$, $C_D$, and $C_E$ show the channels extracted in the Block-B, -C, -D, and

–E, respectively. The '*merge*' function is used for the concatenation of these extracted channels. $\mathbf{u}_d$ shows the number of neurons in Eq. (5).

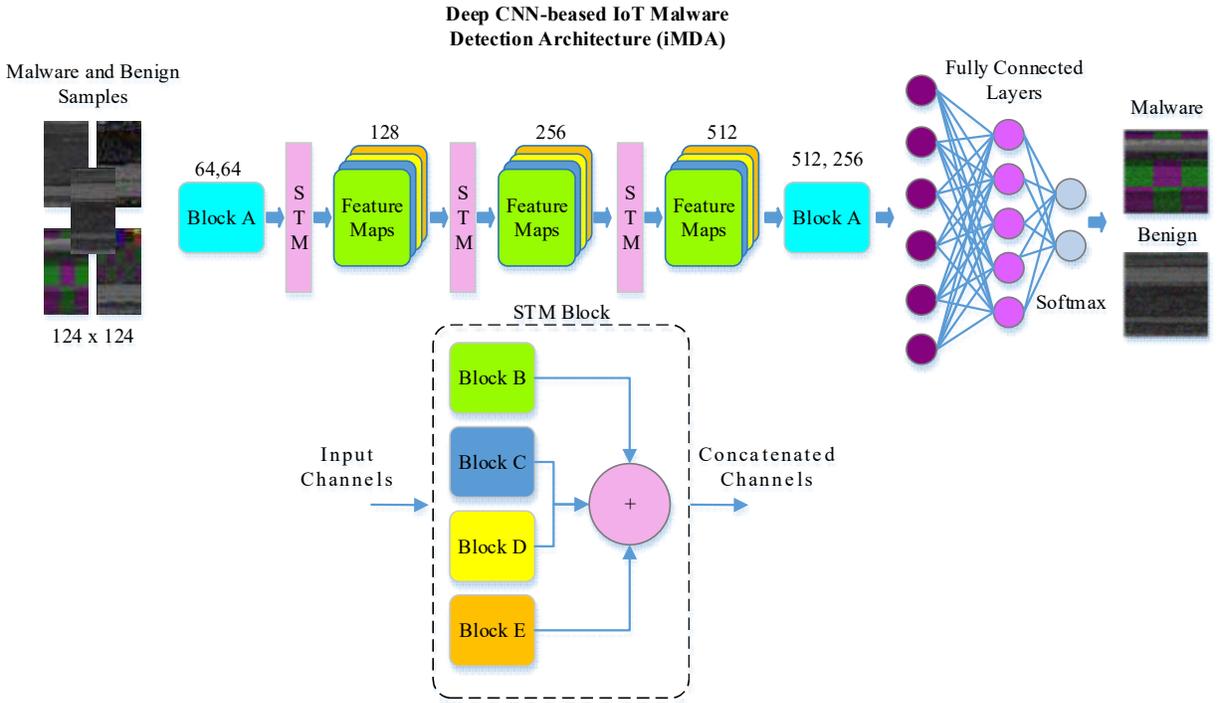

**Figure 4.** (a) The proposed IoT Malware Detection Architecture (b) STM Block (c) Details of Blocks used in the architecture

### 3.4. Implementation of Customized existing CNNs

CNN architectures AlexNet, VGG16, inceptionv3, VGG19, Resnet50, Shufflenet, DenseNet201, Xception, and GoolgeNet are selected for the fair comparison with the proposed architecture. To achieve substantial performance, these models are initially trained on the ImageNet. These trained models are fine-tuned according to the target IoT malware dataset. Then these models are trained and tested using target dataset using 80-20 train-test split.

## 4. EXPERIMENTAL SETUP

### 4.1. Dataset

Linux operating system (OS) is becoming the dominant for IoT devices [31] ], making Linux a prospecting target for the malware developer. Linux uses ELF file format for the deployment of applications or firmware. ELF files are cross plate form in nature and come in two binary formats, packed and unpacked binaries [32]. IOT_Malware dataset used in this study is the image representation of unpacked binary files for malware and benign applications [33]. This dataset is a standard Kaggle benchmark dataset for IoT malware detection challenges. There are 2486 and 14733 image samples for benign and malware application binaries. Visualization of benign and malicious files is shown in Figure 5.

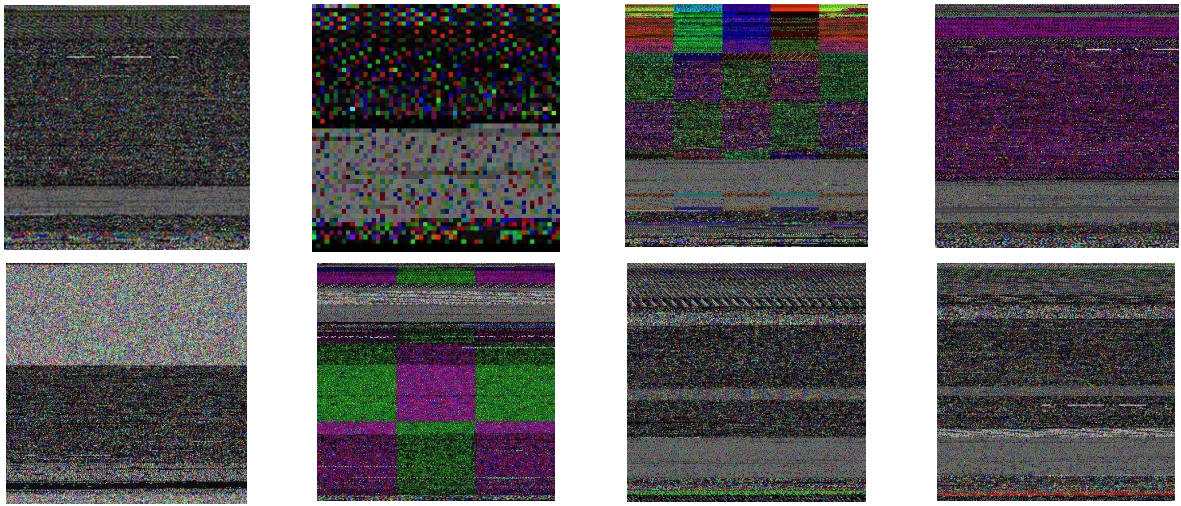

a) Visual Representation of Malware Binaries

`

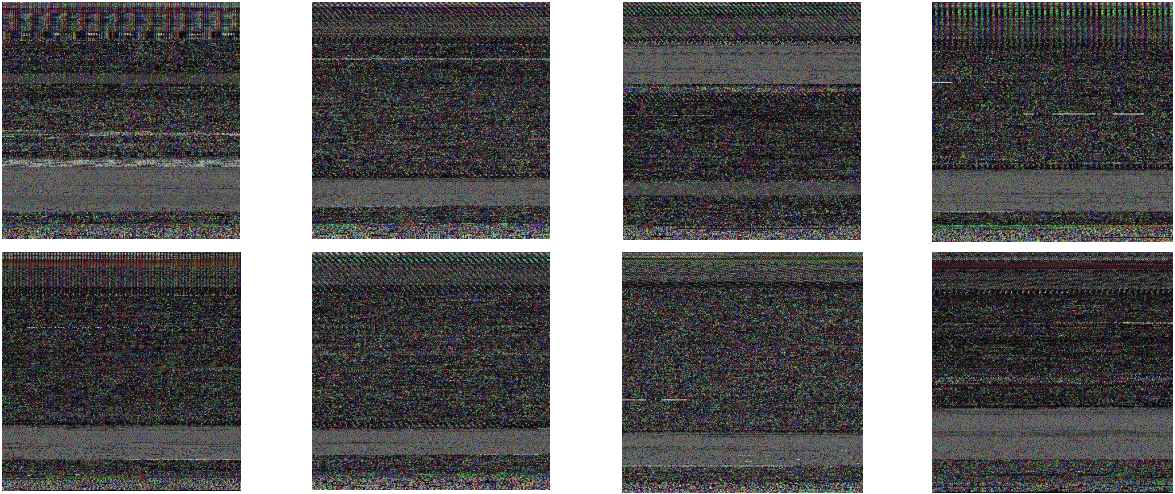

b) Visual Representation of Benign file Binaries

**Figure 5.** Image visualization of (a) Malware and (b)Benign Files

## 4.2. Implementation details

The implementation of the proposed iMDA is simulated using MATLAB-2021a on Nvidia® GTX 1060-T, GPU-enabled Dell Core I i5-7500. It took ~1–2 hours to train a model on the said settings. One epoch took 7–10 minutes on Nvidia-Tesla K-80, while a single IoT malware image took approximately 2 seconds for detection.

## 4.3. Performance Evaluation Metrics

In the current study, we have employed performance evaluation metrics Accuracy, Precision, Recall, F1-Score, and MCC, as shown in Eqs. (6-10). The details of these performance metrics are described in Table 1. AUC-PR and AUC-ROC are also formulated for the proposed model. True Positive (TP), False Positive (FP), True Negative (TN), and False Negative (TN) are also calculated for the performance comparison.

**Table 1** Details of Performance Metrics

| Metric Symbol | Description |
|---|---|
| Acc | Shows Accuracy as % of the total number of Malware detection |
| R | Shows Recall, which is the proportion of correctly identified malware samples and benign samples |
| P | Shows Precision, a ratio of correctly detected malware samples to the total malware sample |
| F1-Score | F1-Score is the harmonic mean of P and R. |
| AUC-PR | Quantifies the area under Precision and Recall Curve |
| AUC-ROC | Quantifies the area under Receiver Operating Characteristic curve |
| MCC | Mathews Correlation Coefficient |
| TP | Correctly Identified Malware Files |

`

| | |
|---|---|
| TN | Correctly Identified Benign Files |
| FP | Incorrectly Identified Malware Files |
| FN | Incorrectly Identified Benign Files |

$$\text{Acc} = \frac{\text{Predicted Malware Samples + Predicted Benign Samples}}{\text{Total Samples}} \times 100 \quad (6)$$

$$\text{MCC} = \frac{(TP*TN)-(FP*FN)}{\sqrt{(TP+FP)*(FP+FN)*(TN+FP)*(TN+FN)}} \quad (7)$$

$$P = \frac{\text{Predicted Malware Samples}}{\text{Predicted Malware Samples + Incorrectly Predicted Malware Samples}} \times 100 \quad (8)$$

$$R = \frac{\text{Predicted Malware Samples}}{\text{Total Malware Samples}} \times 100 \quad (9)$$

$$F1 - \text{Score} = 2 \times \frac{P \times R}{P+R} \quad (10)$$

Several statistical measures are used as performance metrics for binary classification using four quadrants confusion matrix, i.e., TP, TN, FP, and FN. These metrics are selected according to the problem under investigation. There is no agreed-upon performance metrics for two or multi-class problem. The severity of the problem gives direction towards the selection of performance metrics. For an imbalanced dataset, some performance metrics show over-optimistic results. The Matthews correlation coefficient (MCC) is considered an attested statistical measure. It gives a high score for prediction only if all four quadrants are proportionally high for both positive and negative classes [34].

## 5. RESULTS AND DISCUSSION

### 5.1. Performance analysis of the proposed iMDA

The performance of the proposed iMDA is assessed on a standard IoT Malware dataset. F1-Score and MCC are considered standards for performance evaluation for an imbalanced dataset. F1-Score and MCC are used for assigning weightage to both the precision and sensitivity. The proposed architecture converged smoothly and reached the optimal value quickly, as shown in the training plots of the model. Misclassification occurred is due to the intrinsic code similarity between the malicious and benign files. This similarity refers to the identical attack pattern in the malware images. This phenomenon occurred substantially with implementing other CNN models for malware detection. The iMDA is carried out using data augmentation techniques that improve the generalization and robustness of the trained model during testing.

`

## 5.2. Performance comparison with existing CNNs

The performance of the IoT malware detection architecture, iMDA, is also compared with existing models, AlexNet, VGG16, inceptionv3, VGG19, Resnet50, Shufflenet, DenseNet201, Xception, and GoolgeNet. Improved performance is shown in Table 2. The proposed iMDA better explored textural variation in the malware images by systematically using region and boundary information through the Avg and Max-pooling operations. Channel split-transform-merge technique helped to extract the features at different granularity. Incorporating the concepts mentioned earlier in CNN improved the performance of the proposed architecture over the existing models. This study reported the significance of performance using deep learning architecture and quantified it using MCC, F1-Score, AUC-ROC, Accuracy, Precision, and Recall.

**Table 2** Comparison of Proposed Framework with the existing CNN Models

| Models | Accuracy % | F1-Score | Precision | MCC | Recall | AUC-PR | AUC-ROC |
|---|---|---|---|---|---|---|---|
| AlexNet | 92.86 | 0.6807 | 0.9960 | 0.5874 | 0.5171 | 0.9041 | 0.9685 |
| VGG16 | 94.72 | 0.9146 | 0.9552 | 0.839 | 0.8772 | 0.9321 | 0.9816 |
| Inceptionv3 | 94.89 | 0.8055 | 0.9920 | 0.7091 | 0.6780 | 0.8972 | 0.9860 |
| VGG19 | 95.38 | 0.8353 | 0.9902 | 0.7429 | 0.7223 | 0.9088 | 0.9739 |
| Resnet50 | 95.62 | 0.8282 | 0.9971 | 0.7379 | 0.7082 | 0.9432 | 0.9848 |
| Shufflenet | 95.93 | 0.8491 | 0.9949 | 0.7621 | 0.7404 | 0.9541 | 0.9901 |
| DenseNet201 | 96.17 | 0.8685 | 0.9917 | 0.7856 | 0.7726 | 0.9471 | 0.9884 |
| Xception | 96.57 | 0.9342 | 0.9737 | 0.8651 | 0.9074 | 0.9527 | 0.9882 |
| GoolgeNet | 96.72 | 0.8934 | 0.9917 | 0.8195 | 0.8128 | 0.9469 | 0.9881 |
| **Proposed iMDA** | **97.93** | **0.9394** | **0.9864** | **0.8796** | **0.8873** | **0.9731** | **0.9938** |

## 5.3. Detection capability of the proposed iMDA

The effectiveness of a malware detection framework is mainly assessed through precision rate and detection rate. The accurate detection of infused malware in a system is the first parameter to secure and control the spread. Comparison of detection performance of our proposed model using F1-Score, Accuracy, and MCC with the existing model is shown in Figure 6. In contrast, customized existing CNNs are compared and found that few models showed considerably good precision with poor recall. Our proposed model leveraged the difference by comparing F1-Score, the harmonic mean of both parameters. Minimum and maximum performance gain against the existing CNN models are shown in Figure 7. Results of the proposed iMDA are summarised in Table 3.

`

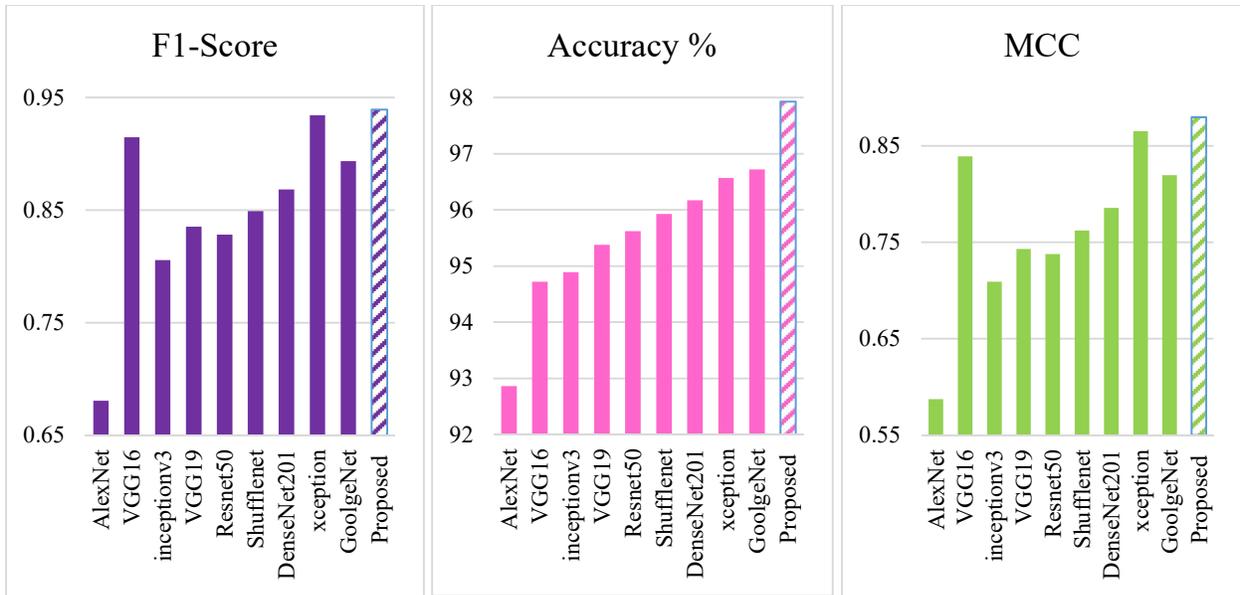

**Figure 6.** F1-Score, Accuracy, and MCC Comparison

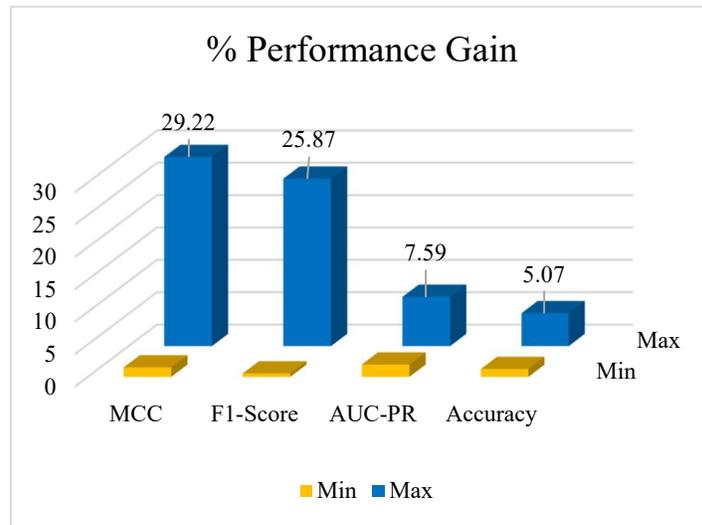

**Figure 7.** Minimum and Maximum Performance Gain of Proposed Framework

**Table 3** Performance of the proposed model

| Net | Accuracy % | F1-Score | Precision | MCC | Recall | AUC-PR | AUC-ROC |
|---|---|---|---|---|---|---|---|
| Proposed iMDA | 97.93 | 0.9394 | 0.9864 | 0.8796 | 0.8873 | 0.9689 | 0.9938 |

## 5.4. *Feature space-based analysis of the proposed iMDA*

The decision-making of the proposed architecture is better analyzed with the help of feature space visualization. Better discrimination factor of the model is associated with the prominent visual

`

features. This distinction helps to improve the learning and lower the variance of the model. The feature space visualization for the principal components of our proposed iMDA is shown in Figure 8. Channel squeezing and channel boosting used in STM blocks helped to capture the discriminative features of the IoT malware images at a multi-level. Additionally, STM boosted the reduced prominent feature with the help of channel concatenation. The feature space visualization for the proposed iMDA showed an improvement in identifying the distinct and diverse features and hence improved the detection of the IoT malware files.

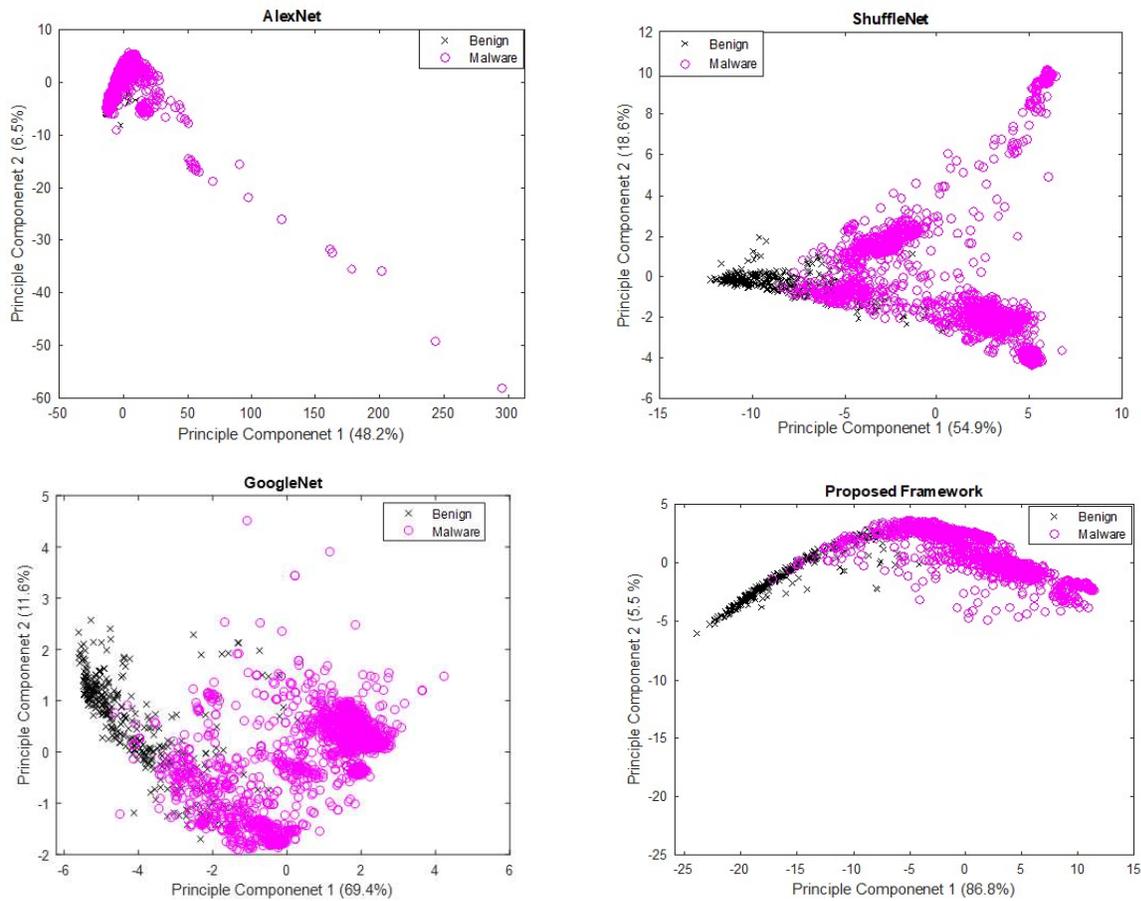

**Figure 8.** Feature Space-based performance comparisons

## 5.5. *AUC-ROC and AUC-PR based analysis*

The optimal performance of the model is also best understood by the ROC and PR plots. These plots show the bifurcation capability of the models at an optimal threshold value. Our proposed iMDA showed high sensitivity along with a decreased false positive rate.

`

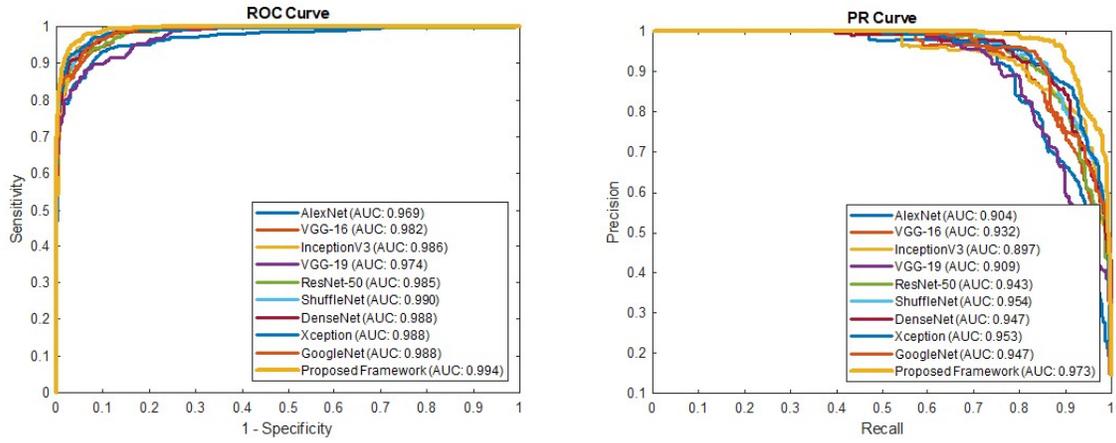

**Figure 9.** AUC-ROC and AUC-PR

## 6. CONCLUSION

Analysis of malware in IoT is an early line of defense in securing this world of connected devices from cyberattacks. Malware analysis help to identify and designate the malicious code segments hidden in the legitimate files. This malicious code snippet is present according to the malware signature or obfuscated otherwise. The obfuscation techniques hide the malicious code lines from pattern/signature matching. These lines may be distributed or intermixed with the legitimate line over the complete file. The proposed iMDA exploits the concepts of dilated convolutional operations in STM blocks, channel squeezing, and boosting to extract minor contrast and texture variation. Moreover, dilated convolutional operations followed by alternative application of the avg- and max-pooling in multi-path help learn local and global structural patterns at different granule levels. The proposed iMDA outperformed existing CNN and achieved the best result for the Accuracy (97.33%), MCC (.8796), F1-Score (93.94), AUC-ROC (.9938), and AUC-PR (.9689). In the future, the proposed iMDA may be extended for the android-based malware detection and IoT Elf files compositely.

`

*Code availability*

All the scripts that are developed for the simulations are available from the corresponding author on reasonable request.

*Declaration of interests*

The authors declare that they have no known competing financial interests or personal relationships that could have appeared to influence the work reported in this paper.

*Declaration of Competing Interest*

The authors declare no conflict of interest.

*Acknowledgments*

This work was conducted with the support of the PAEC program. As well as, we thank Pattern Recognition Lab (PR-Lab) and Pakistan Institute of Engineering, and Applied Sciences (PIEAS), for providing necessary computational resources and a healthy research environment.

`

`

`

`